# Exploiting Map Topology Knowledge for Context-predictive Multi-interface Car-to-cloud Communication

**Benjamin Sliwa, Johannes Pillmann, Maximilian Klaß and Christian Wietfeld**
Communication Networks Institute
TU Dortmund University, 44227 Dortmund, Germany
e-mail: {Benjamin.Sliwa, Johannes.Pillmann, Maximilian.Klass, Christian.Wietfeld}@tu-dortmund.de

***Abstract*—While the automotive industry is currently facing a contest among different communication technologies and paradigms about predominance in the connected vehicles sector, the diversity of the various application requirements makes it unlikely that a single technology will be able to fulfill all given demands. Instead, the joint usage of multiple communication technologies seems to be a promising candidate that allows benefiting from characteristical strengths (e.g., using low latency direct communication for safety-related messaging). Consequently, dynamic network interface selection has become a field of scientific interest. In this paper, we present a cross-layer approach for context-aware transmission of vehicular sensor data that exploits mobility control knowledge for scheduling the transmission time with respect to the anticipated channel conditions for the corresponding communication technology. The proposed multi-interface transmission scheme is evaluated in a comprehensive simulation study, where it is able to achieve significant improvements in data rate and reliability.**

## I. Introduction

Within the ongoing struggle between IEEE 802.11p and 5G about which technology will win the race for becoming the enabler for autonomous driving and a key factor for smart city-based Intelligent Transportation Systems (ITSs) [1], more and more research works address the joint-usage of both technologies in order to exploit specific strengths and compensate corresponding shortcomings [2]. While the main use-case for WiFi-based approaches is near-field safety-related communication such as the distribution of Cooperative Awareness Messages (CAMs), an additional cellular connection will serve for uplink car-to-cloud communication using the vehicle as a mobile sensor, e.g. for predictive maintenance, traffic monitoring and intelligent traffic control (cf. Fig. 1).

In earlier work [3], we presented the Channel-aware Transmission (CAT) scheme that takes into account the channel quality based on the Signal-to-noise ratio (SNR) of the Long Term Evolution (LTE) downlink signal for determining efficient transmission times in order to avoid resource-intense transmissions during unfavorable channel situations. Another significant boost in the transmission efficiency was achieved by machine learning based throughput prediction in [4] using all available LTE downlink indicators as well as knowledge about the payload size. While the previous work only addressed pure cellular data transmission via a single LTE link, the applicability of the context-aware transmission approach for multi-interface applications with heterogeneous communication technologies is evaluated in this work. Furthermore, the availability of mobility information obtained from the navigation system is exploited to integrate knowledge about the surrounding buildings and the anticipated trajectory of the vehicle into the transmission decision process and take the shadowing effects into account for the channel assessment.

The paper is structured as follows. After giving an overview about relevant state-of-the-art approaches, the system model of the solution approach is presented and the analytical model of the transmission scheme is discussed. Afterwards, the setup for the simulative evaluation is introduced and finally, detailed evaluations focusing on specific indicators and performance measures are discussed.

## II. Related Work

The research fields related to this work are classified by the topics *context-aware communication* and *multi-interface data transmission*. The anticipatory communication paradigm [5] has been proposed for paying attention to the channel dynamics mainly related to the mobility behavior of the vehicles. Mobility knowledge and context information is ex-ploited to optimize decision processes within the communication systems themselves. In [6], the problem of building-related shadowing in vehicular communication scenarios is discussed and addressed by dynamic beaconing strategies. A comprehensive survey about interworking of IEEE 802.11p and cellular network technologies for Vehicle-to-everything (V2X) applications is provided in [7] and a simulation-based study of various application requirements for both technologies is presented in [8]. The authors of [9] bring together the aforementioned research fields for infrastructure-assisted selection of the currently best radio access technology. Within a simulative performance evaluation, they achieve significant benefits for downloading data from a server before a defined deadline has expired.

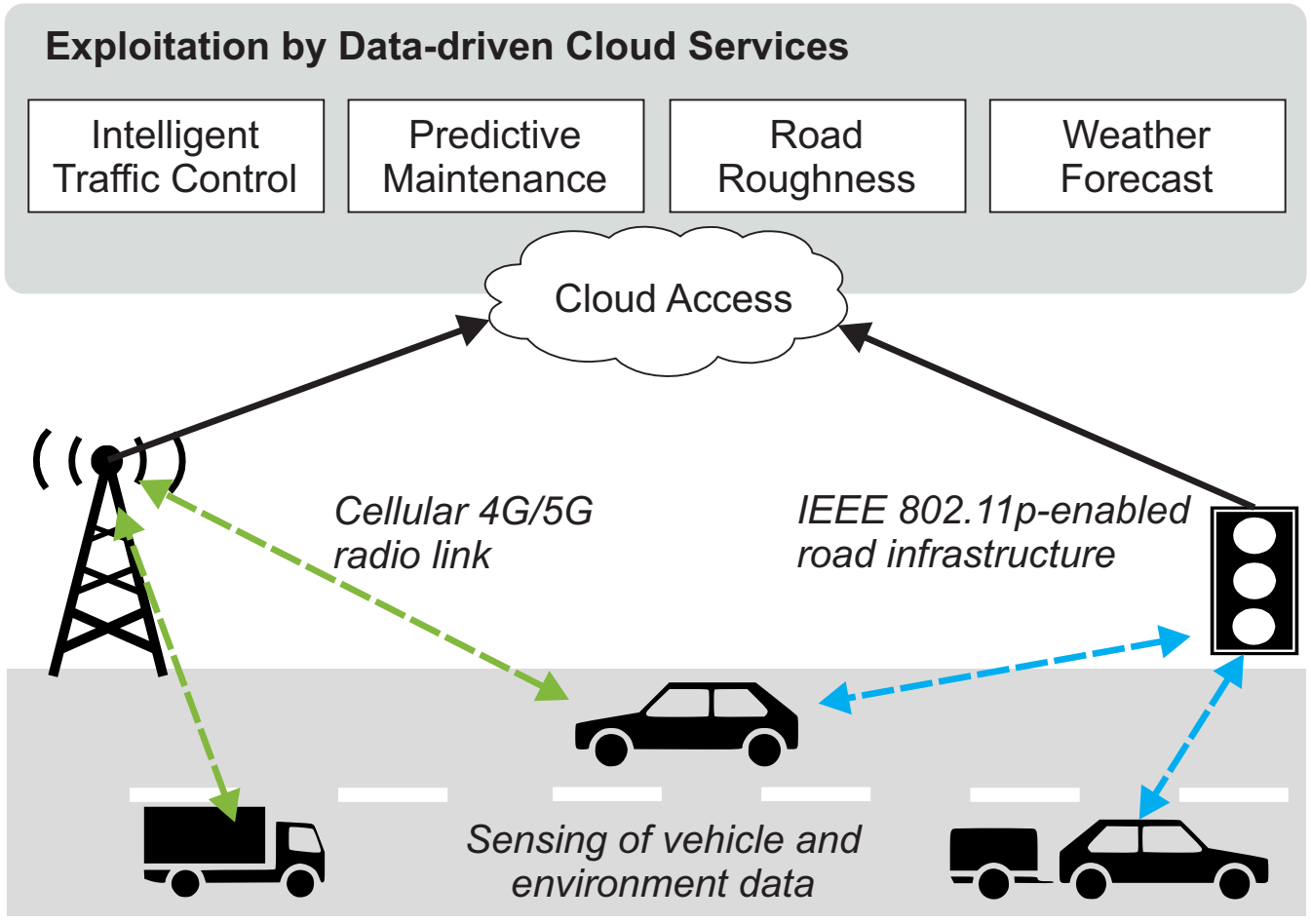

Fig. 1. Example application scenario: Vehicles collect sensor data which is then transmitted to the cloud via the radio link with the highest available network quality.

A hierarchical method with a centralized server for interface selection is proposed in [10]. The network interface for opportunistic data transmissions is selected with respect to the communication region based on previous measurements that are maintained in a database. In order to pay attention to the time-variance of the transmission channel, vehicles have to periodically download the most recent decision strategies. Similarly, [11] proposes interface selection based on Quality of Service (QoS) requirements for enabling handovers between different technologies. Another approach for exploiting the presence of multiple network interfaces, that is currently receiving great attention within the vehicular communication community, is the application of network coding methods. In [12], the authors propose a network-coding-assisted scheduling scheme for Vehicle-to-Vehicle (V2V) and Vehicle-to-Infrastructure (V2I) communication. While this specific topic is not further discussed in the context of this work, it will be considered for future extensions of the proposal. In contrast to existing solutions that integrate the infrastructure side into the interface selection process and which are therefore unlikely to be realized in real-world scenarios due to the required changes on the network provider side, the proposed approach works decentralized and only uses locally accessible information.

## III. Simulation-based Solution Approach

In this section, the overall system model as well as the analytical model for the transmission scheme are presented. Vehicles perform uplink transmissions of sensor data to a cloud-server using the proposed context-aware transmission schemes in order to exploit favorable channel conditions for increased data rate and reliability.

The processing steps for context-aware multi-interface data transmission are illustrated in Fig. 2. The generated sensor data is stored in a local buffer until a positive sending decision is made. Mobility information as well as current channel quality measurements provide the transmission context and influence the sending decision using mobility prediction and channel quality estimation. Each host evaluates the obstacle channel model of [13] for the current and predicted position. Thus a sending probability is calculated and evaluated before transmitting the whole data of the buffer. Several system parameters can be changed to influence the sending decision and the interface selection characteristics.

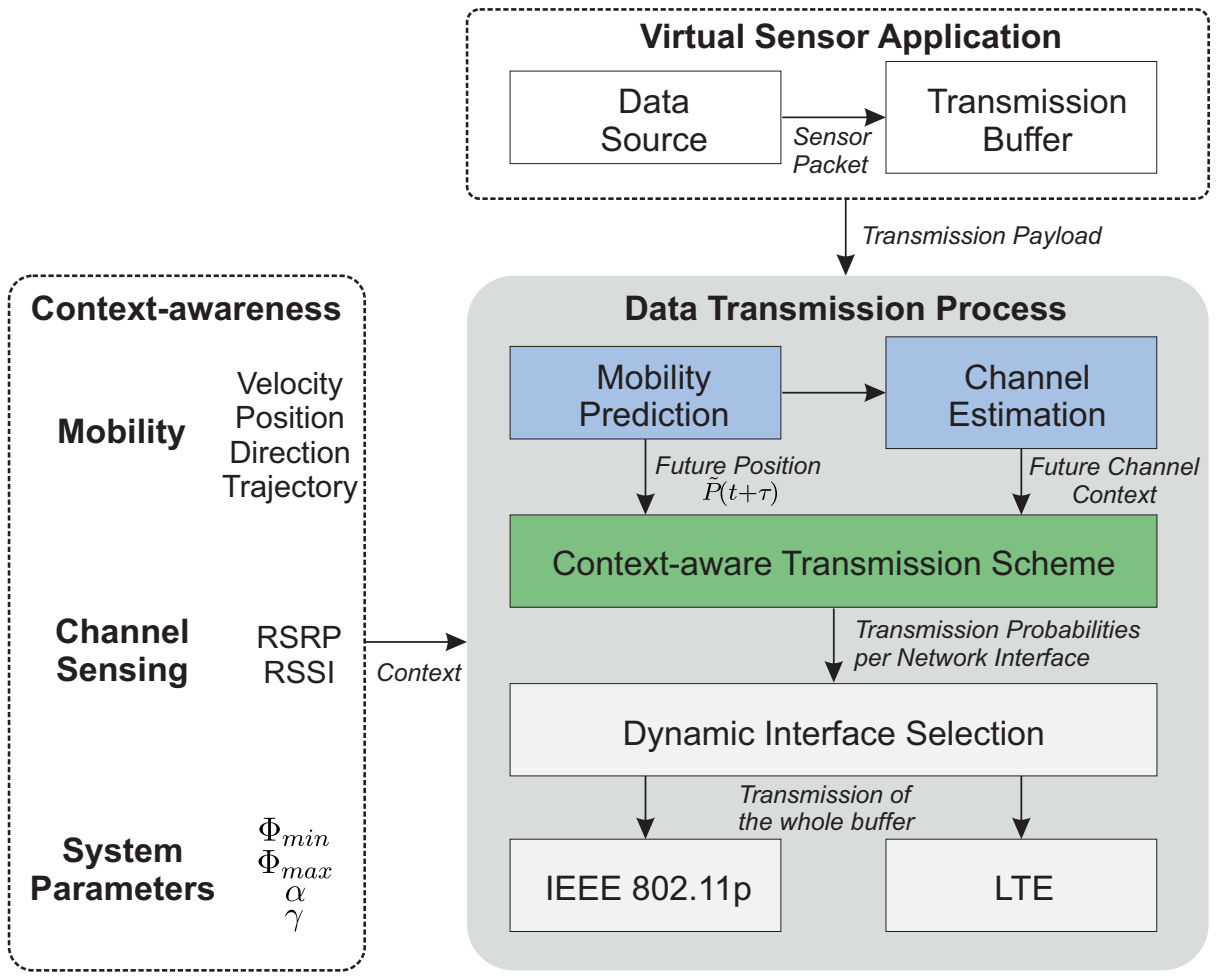

Fig. 2. System model of the context-predictive multi-interface transmission approach.

### A. Mobility Prediction

In the course of this work, three different mobility prediction methods are implemented and evaluated that differ in the utilized mobility information. Fig. 3 gives a brief overview of the concept of mobility prediction and the implemented channel model considering obstacles in the line of sight (LOS). A major goal of the transmission process is to avoid transmissions while buildings are in the LOS, as they have a high impact on the channel quality. Thus, mobility prediction is applied to find a future location with an expected enhanced channel quality. The first algorithm uses simple linear *Extrapolation* to predict the vehicle's future position based on its current speed $v(t)$ and direction for a given prediction horizon $\tau$. The traveled distance $s_{max}$ is calculated as:

$$s_{max} = v(t) \cdot \tau \tag{1}$$

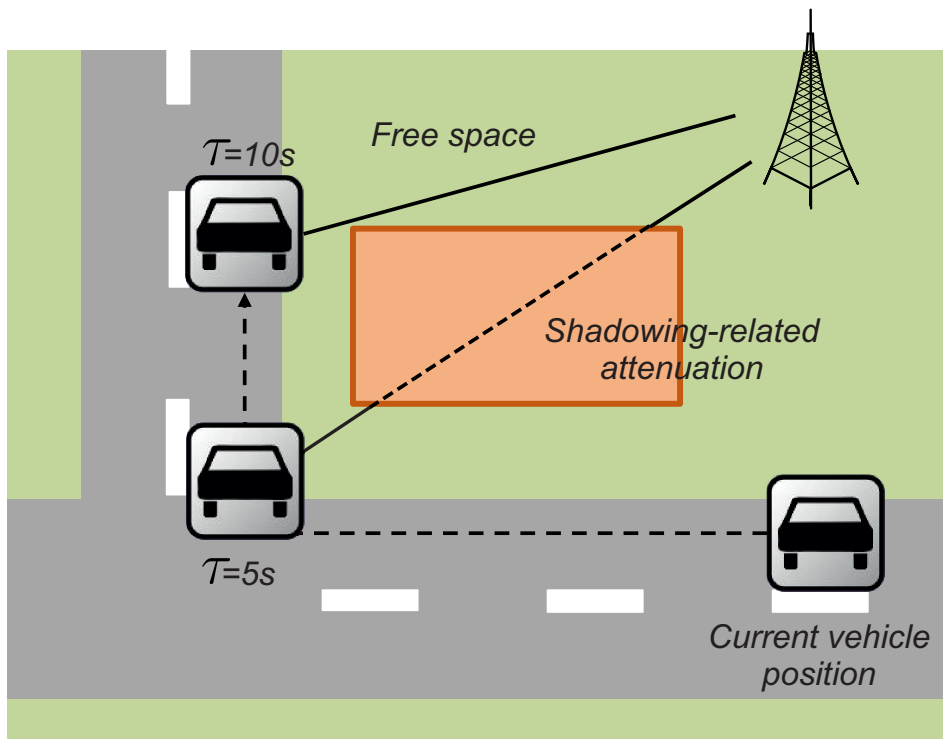

Fig. 3. Mobility prediction with different values for the prediction horizon $\tau$. Here, the data transmission is most likely to be triggered at $\tau = 10s$, after the vehicle has passed the building-related attenuation region.

The second method $Trajectory_{vel}$ uses the calculated value of $s_{max}$ but considers the vehicle's route by accessing the navigation system.

$$s_{max} = \begin{cases} 0.5 \cdot a(t)^2 \cdot \tau + v(t) \cdot \tau & v(t) < v_{max} \\ v_{max} \cdot \tau & v(t) = v_{max} \end{cases} \quad (2)$$

For the third method $Trajectory_{acc}$, the calculation of $s_{max}$ takes the current acceleration $a(t)$ into account using Eq. 2. The acceleration-based term is used until the maximum speed allowed on the road segment is reached. This method aims to provide a better model for the vehicle's behavior at traffic signals and is only applied for positive acceleration values.

Based on the calculated maximum traveling distance the future position of the vehicle is obtained, following the route, if available. Afterwards, for the channel quality estimation, the channel model is evaluated for the predicted position.

### B. Single-interface Transmission Scheme

Fig. 4 shows an example trace of the SNR and compares periodic transmissions (which is performed regardless of the current channel situation) with the original probabilistic CAT as described in [3], which calculates a transmission probability with respect to the measured SNR value. Thus, good channel conditions are chosen more likely for transmitting data and result in an enhanced mean data rate.

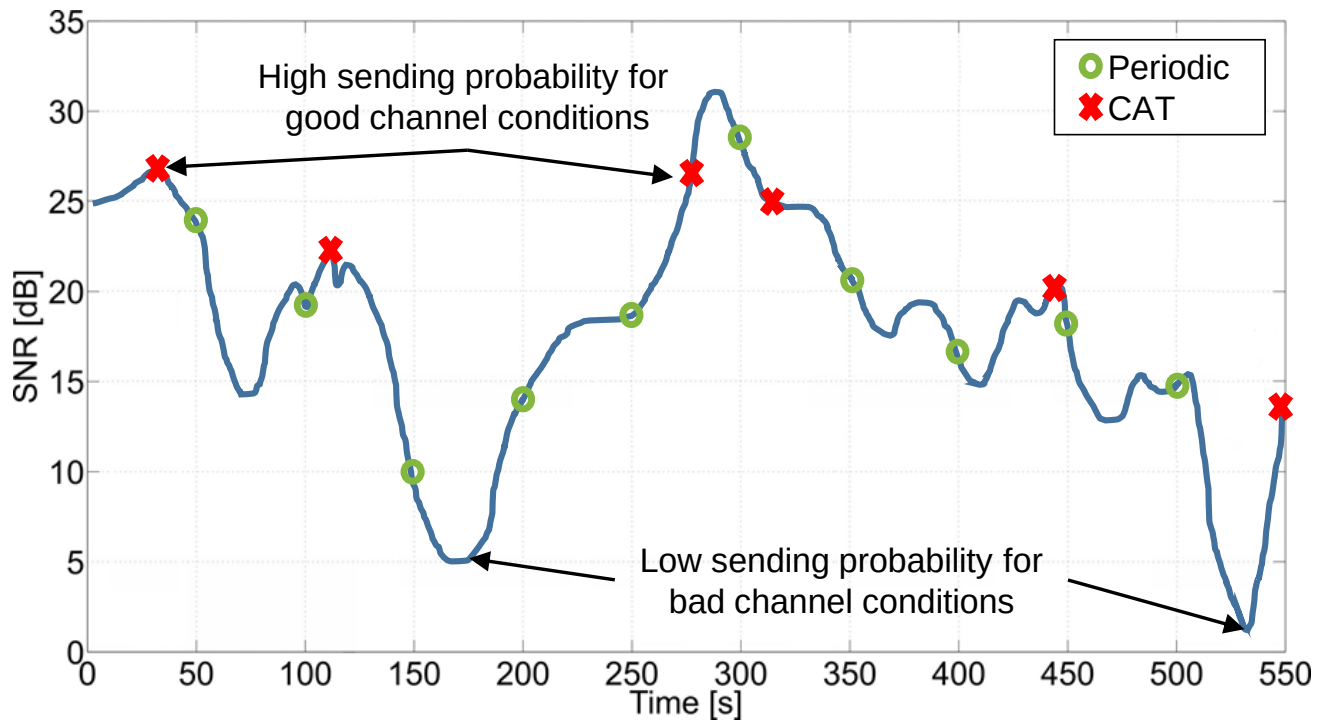

Fig. 4. Example temporal behavior of the CAT [3] transmission scheme in comparison to naive periodic data transmissions that do not take the channel quality into account.

In order to enable multi-interface transmissions with heterogeneous communication technologies, the implicitly SNR-normed metric of the legacy CAT needs to be abstracted to allow the comparison of metrics with different value ranges. Therefore, the abstract metric $\Phi$ is defined by its minimum value $\Phi_{min}$ and maximum value $\Phi_{max}$. With Eq. 3, the current metric value $\Phi(t)$ is normed to its respective value range. The transmission probability $p(t)$, which is calculated according to Eq. 4, is computed in the time interval $[t_{min}, t_{max}]$ in order to avoid too long buffering times and too short data packets. The exponent $\alpha$ defines how much the transmission probability should rely on very high metric values within the respective value range.

$$\Theta(t) = \frac{\Phi(t) - \Phi_{min}}{\Phi_{max} - \Phi_{min}} \quad (3)$$

$$p(t) = \begin{cases} 0 & t - t' \leq t_{\text{min}} \\ \Theta(t)^{\alpha} & t_{\text{min}} \leq t - t' \leq t_{\text{max}} \\ 1 & t - t' \geq t_{\text{max}} \end{cases} \quad (4)$$

For considering the predicted future network state, the extended predictive CAT (pCAT) transmission scheme is introduced with Eq. 5. Depending on the anticipated development of the channel quality, the transmission probability is changed to transmit data early if the channel quality decreases or delay the transmission if it improves. $\gamma_1$ is applied if the channel quality is expected to improve and $\gamma_2$ if it is anticipated to decrease.

$$p(t) = \begin{cases} 0 & t - t' \leq t_{min} \\ \Theta(t)^{\alpha \cdot z_1} & t_{\text{min}} \leq t - t' \leq t_{\text{max}}, \Delta\Phi(t) \geq 0 \\ \Theta(t)^{\frac{\alpha}{z_2}} & t_{\text{min}} \leq t - t' \leq t_{\text{max}}, \Delta\Phi(t) \leq 0 \\ 1 & t - t' \geq t_{\text{max}} \end{cases} \quad (5)$$

$$z_1 = \max\left(\Delta\Phi(t) \cdot (1 - \Theta(t)) \cdot \gamma_1, 1\right) \quad (6)$$

$$z_2 = \max\left(\left|\Delta\Phi(t) \cdot \Theta(t) \cdot \gamma_2\right|, 1\right) \quad (7)$$

$$\Delta\Phi(t) = \overline{\Phi}\left(t, t + \tau\right) - \Phi(t) \quad (8)$$

### C. Multi-interface Transmission Scheme

The proposed multi-interface approach computes the transmission probabilities for every network interface of the vehicle according to Eq. 9. The overall transmission probability is the maximum value of the different individual interfaces. Thereupon, the overall probability is evaluated to send all data in the buffer. As the proposed algorithm is working probabilistically, the selection of the network interface itself is independent of the current transmission scheme, which might be CAT or pCAT.

$$p_{MI} = max\left\{p_i(\Phi(t))\right\} \quad \forall \ \Phi(t) \quad (9)$$

For the following evaluations, two different metrics are defined: $\Phi_{RSRP}$ is based on the Reference Signal Received Power (RSRP) of the LTE link and $\Phi_{RSSI}$ characterizes the Received Signal Strength Indicator (RSSI) of the IEEE 802.11p signal.

## IV. Methodological Setup

The methodological simulation setup is illustrated in Fig. 5 and consists of the third party extension frameworks SimuLTE [14] and Vehicles in Network Simulation (Veins) [15] that use the OMNeT++ [16] simulation environment. In addition, the traffic simulator Simulation of Urban Mobility (SUMO) [17] is used to model the vehicular mobility and to provide the building locations.

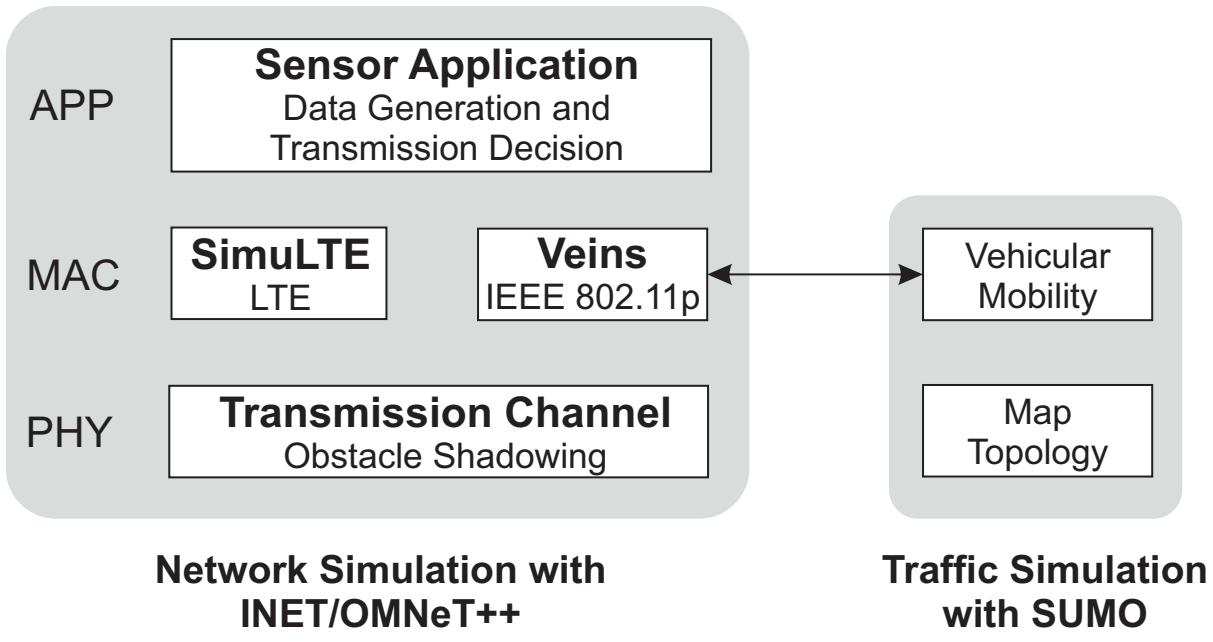

Fig. 5. Coupling of mobility and communication simulator. Frameworks for LTE and IEEE 802.11p are coupled within OMNeT++ for communication simulation.

Although OMNeT++ itself is built upon a modular approach, various changes on the different layers were necessary as the extension frameworks are not designed for being used in a multi-interface use-case and Veins has a strong dependency among the different layers. Similarly, as SimuLTE relies on its own dedicated channel models, the physical layer was brought together with the Veins channel model in order to integrate shadowing effects and establish consistency between the different frameworks.

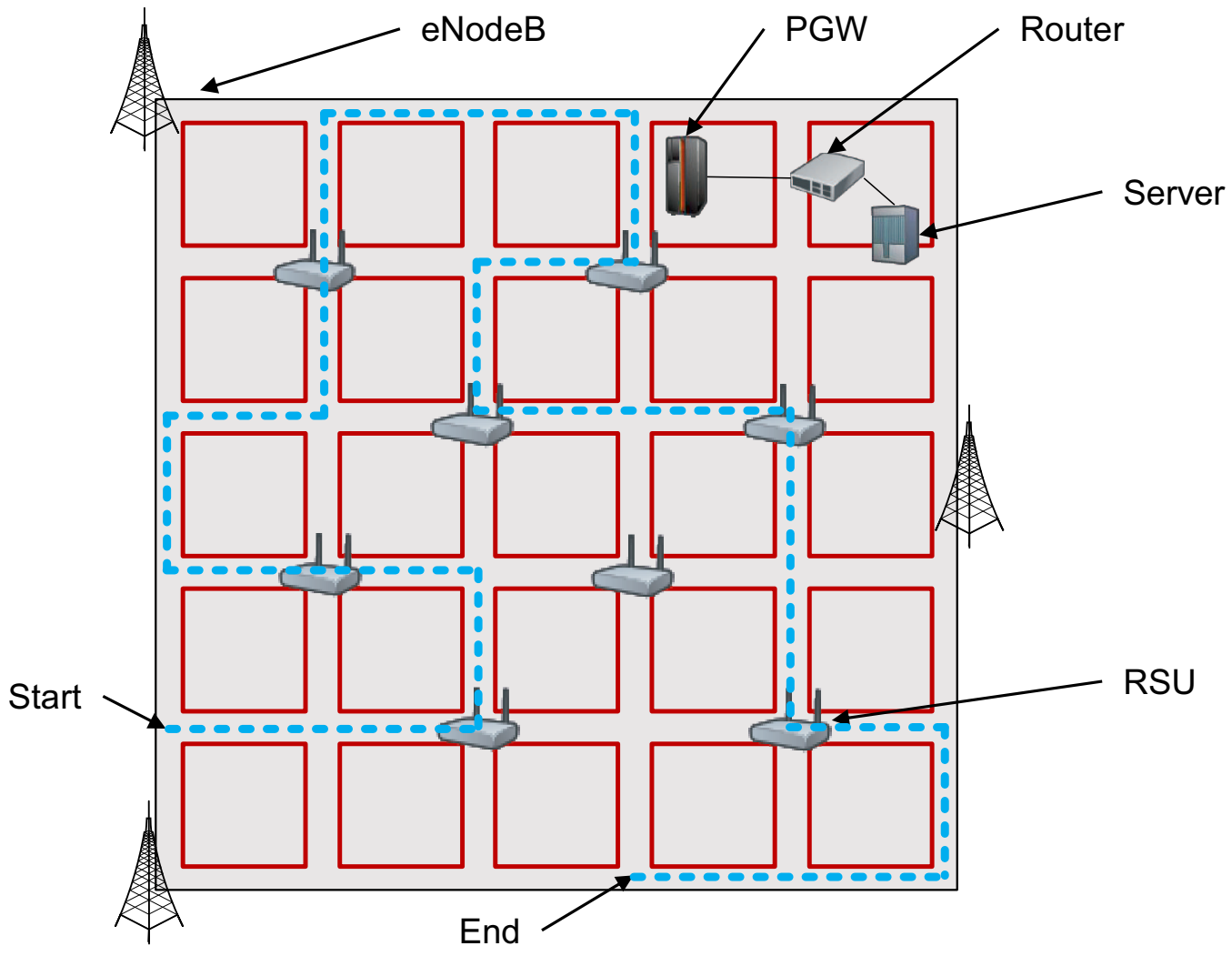

Fig. 6. Generic Manhattan Grid scenario with three LTE cells and eight RSUs and the route of an example vehicle.

Fig. 6 shows the map of the simulation scenario. From all present vehicles, a percentage of $r$ are randomly chosen to be equipped with communication capabilities, the others act as interference traffic. Each vehicle performs and grid-based *Random Walk* and sends sensor data to the cloud-server according to the transmission scheme. There are three LTE cells and eight RSUs in the simulation playground. Fig. 6 shows the route of an exemplary vehicle. In the following section, the temporal behavior of this particular vehicle is used to state out characteristical effects occurring in mobility and channel prediction.

The relevant parameters for the reference scenario are defined in Tab. I.

TABLE I
SIMULATION PARAMETERS

| Parameter | Value |
|---|---|
| LTE carrier frequency | 1800 Mhz |
| eNode B transmission power | 33 dBm |
| eNode B antenna | Omnidirectional |
| IEEE 802.11p carrier frequency | 5.89 GHz |
| Sensor frequency | 1 Hz |
| Sensor payload | 10 kByte |
| Transmission interval $\Delta t$ | 15 s |
| Minimum buffering duration $t_{min}$ | 10 s |
| Maximum buffering duration $t_{max}$ | 60 s |
| Playground size | 1000 m $\times$ 1000 m |
| Number of vehicles | 150 |
| Penetration rate $r$ | 10 % |
| Attenuation per cut $\beta$ | 2 dB |
| $\Phi_{\text{RSRP}}$ [min,max] | [-140 dBm, -50 dBm] |
| $\Phi_{\text{RSSI}}$ [min,max] | [-89 dBm, -50 dBm] |
| $\alpha$ | 8 |
| $\gamma_1$ | 3 |
| $\gamma_2$ | 0.5 |

## V. RESULTS OF THE SIMULATIVE PERFORMANCE EVALUATION

This section evaluates the three different mobility prediction algorithms as well as the resulting channel quality prediction. The results of the simulation runs for an LTE, 802.11p and a multi-interface scenario are presented and discussed.

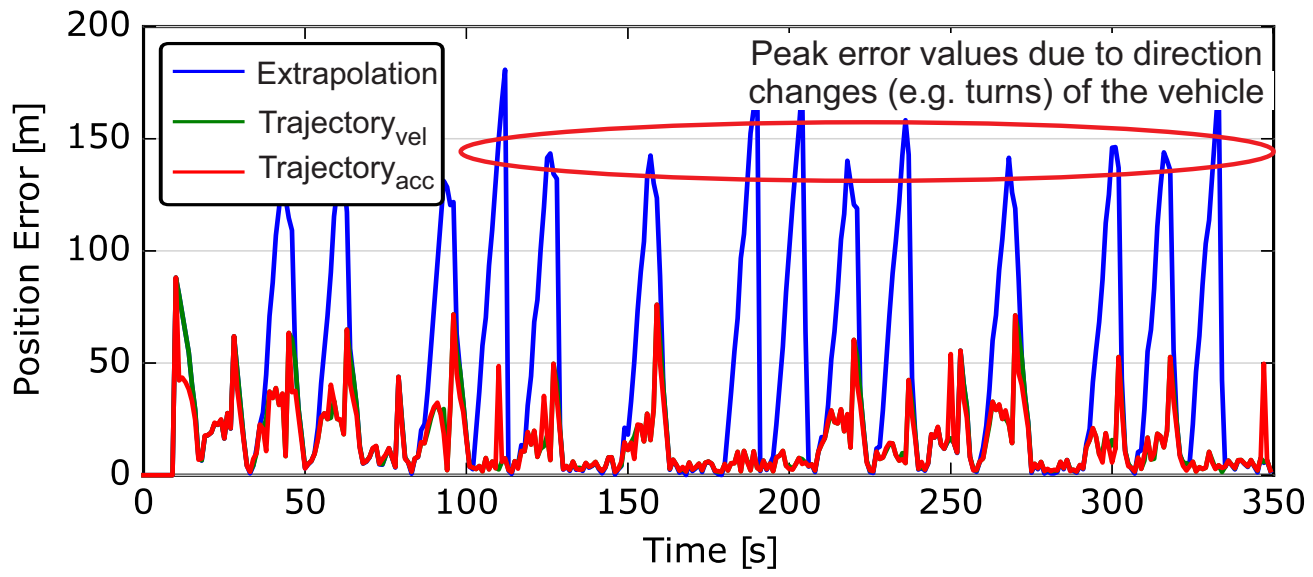

Fig. 7. Temporal behavior of the absolute location prediction error with prediction horizon $\tau = 10\ s$ for a single vehicle.

Fig. 7 shows the absolute location prediction error for the aforementioned vehicle for short-term predictions with $\tau = 10s$. The extrapolation-based algorithm performs worst, as it does not consider turns, resulting in the peaks with high position errors. The trajectory-aware approaches perform significantly better. Additionally, considering the acceleration further reduces the resulting error after the car has stopped or slowed down at intersections and traffic signals.

Fig. 8 shows the statistical properties for all considered mobility prediction schemes with $\tau = 10\ s$, $\tau = 30\ s$ and $\tau = 60\ s$. The achievable error range is mainly influenced by the vehicle's current speed value and its respective time-stability. Interactions with other traffic participants and traffic signals have a major impact on the resulting accuracy as they are not considered by the mobility prediction.

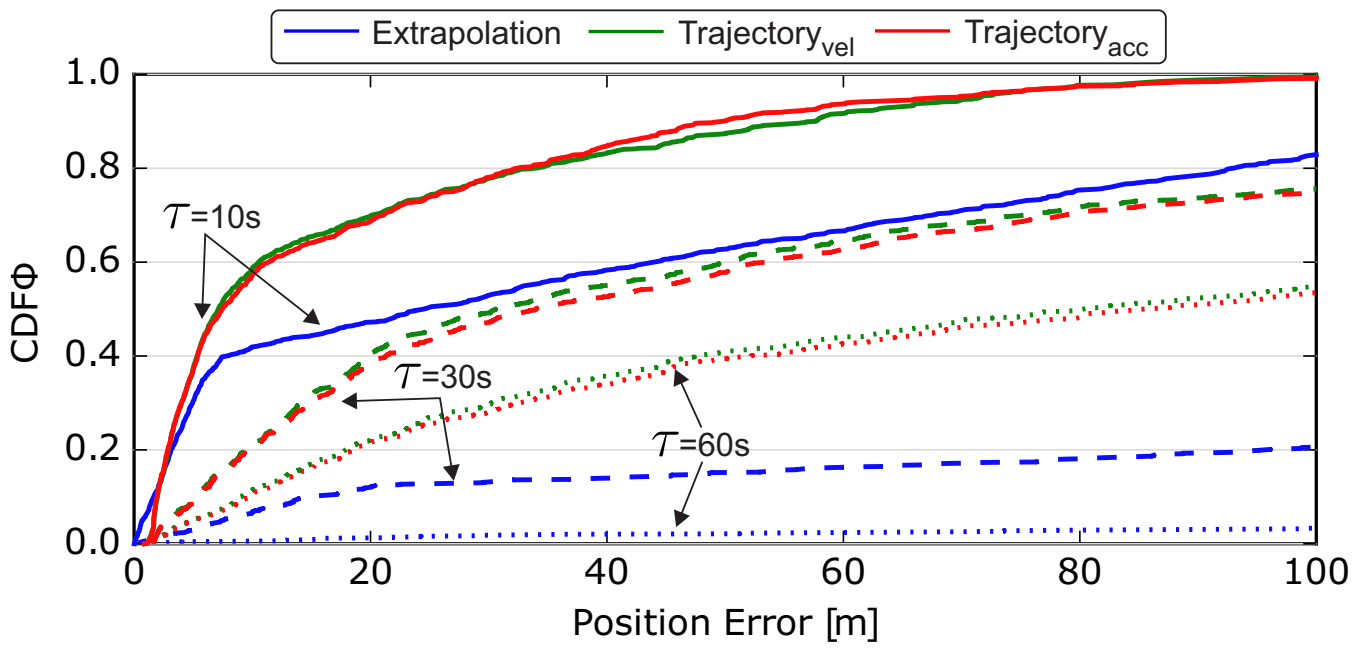

Fig. 8. Accuracy of three different mobility prediction methods evaluated for for all vehicles with prediction horizons of 10 s, 30 s and 60 s.

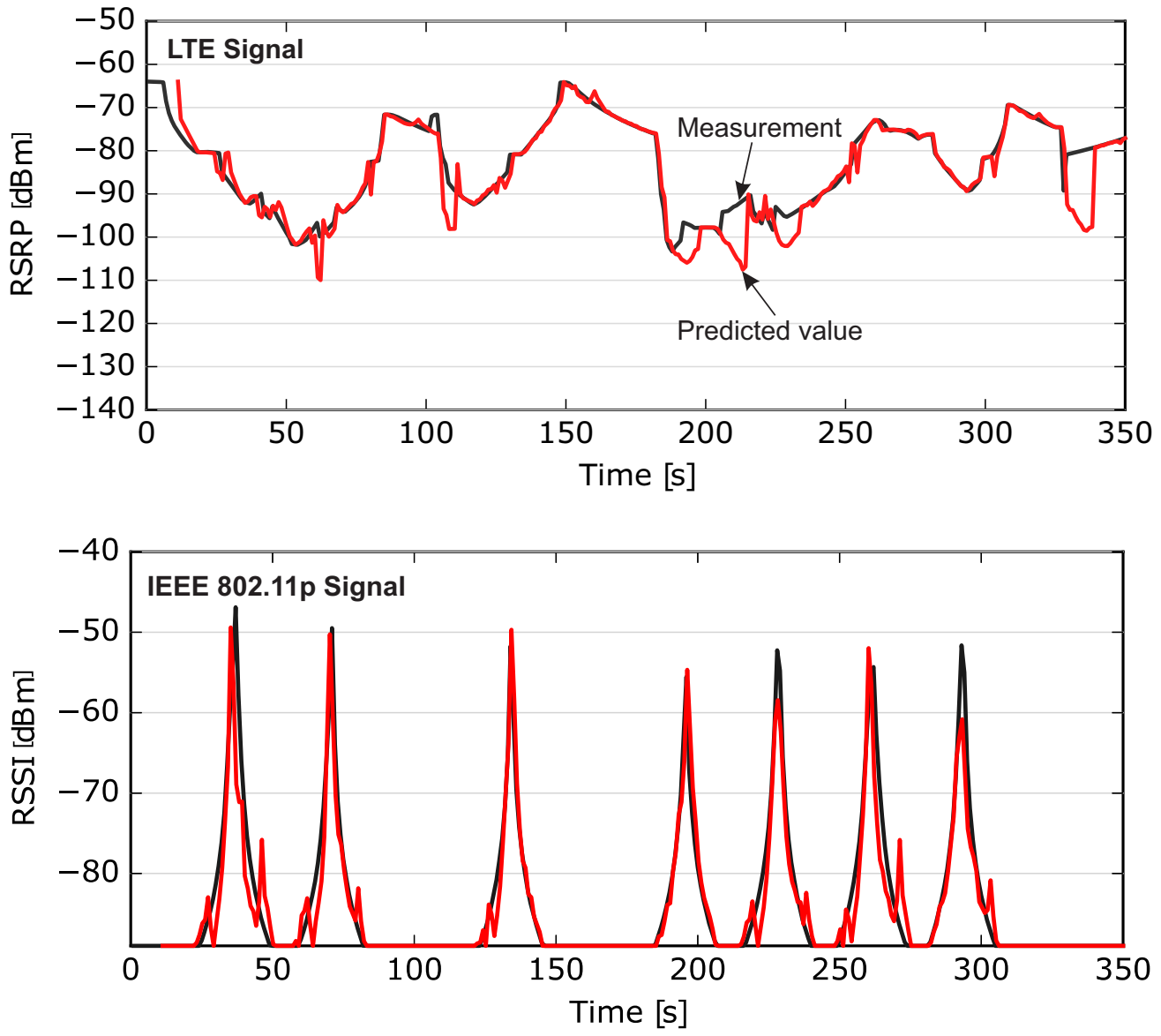

Fig. 9. Temporal behavior of the RSRP and RSSI estimation based on the mobility prediction with acceleration of one example car. The black graph represents the actual measurements, whereas the red graph represents the estimated values.

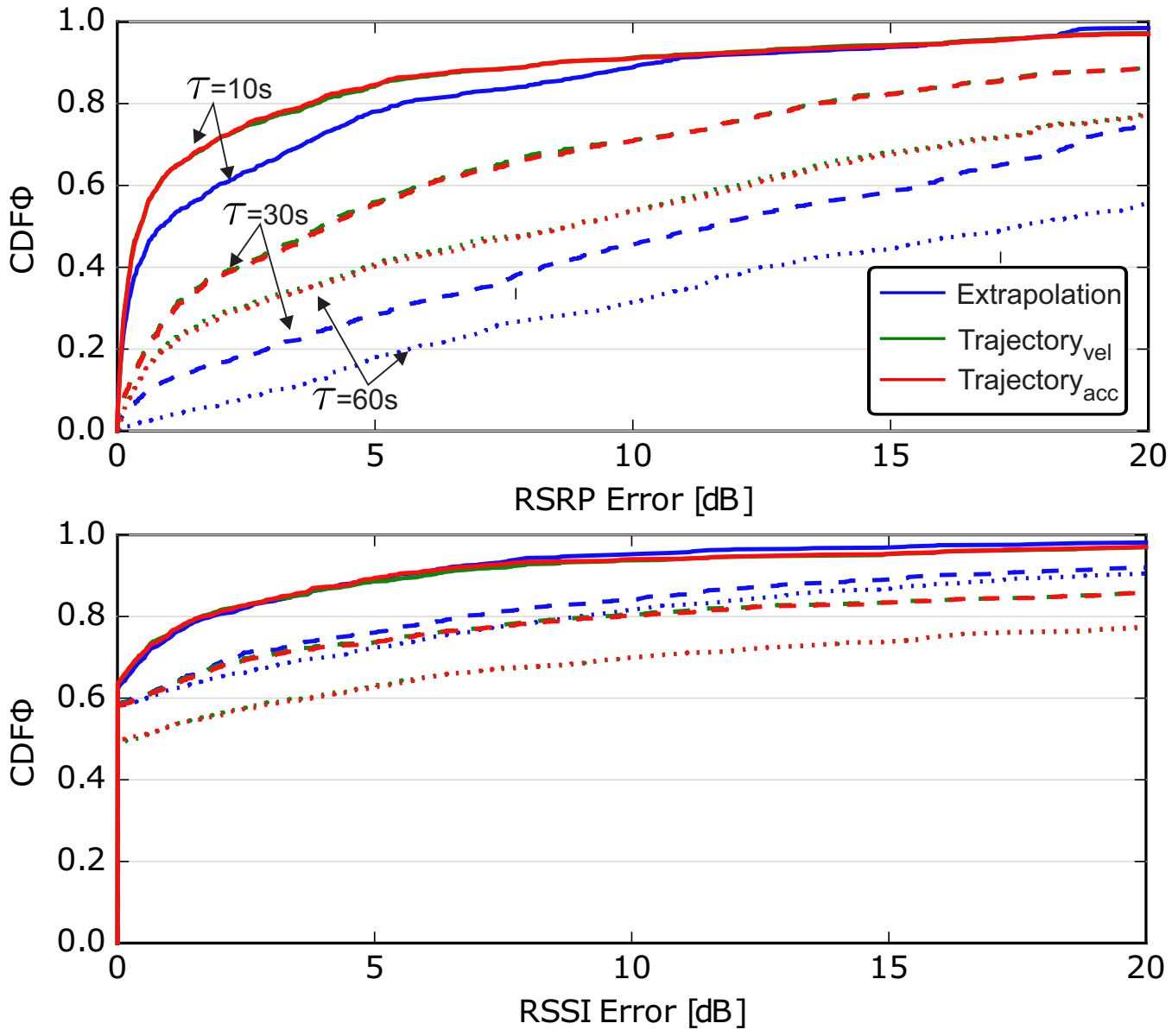

Fig. 10. Accuracy of the RSRP and RSSI estimation for the different mobility prediction methods and prediction horizons.

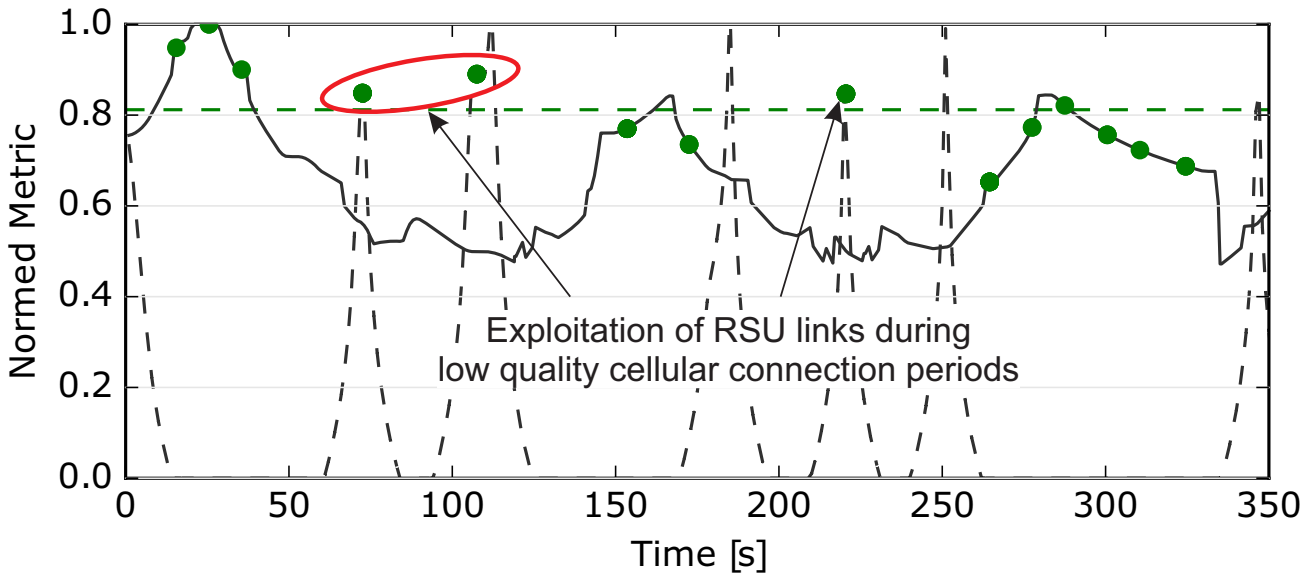

Fig. 11. Temporal behavior for a single vehicle using multiple interfaces for sensor data transmission. The dashed green line indicates the mean value of the metric for a sending decision.

Fig. 9 shows the measured values for RSRP and RSSI for the given route in Fig. 6. Due to the limited range of the RSUs, connection losses occur between the connectivity peaks. As handovers are not considered by the prediction process, the RSRP estimation differs from the measured values resulting in small peaks.

The accuracy of the channel metric prediction mainly relies on the quality of the mobility prediction, as it can be seen in Fig. 10. A smaller prediction horizon reduces the dependency on an accurate speed value and results in a lower prediction error.

Fig. 11 illustrates the temporal behavior of the considered radio links as well as the resulting transmission scheme. The sending algorithm chooses the interface with the best channel quality and thus avoids connectivity valleys of each technology by choosing the other.

The resulting overall values for the considered key performance indicators goodput, sensor data age (which is mainly determined by the duration a sensor packet is stored in the transmission buffer) and packet delivery ratio are shown in Fig. 12, which compares the application-layer performance of the solely application of a single communication technology with the multi-interface approach. The results show that the achievable data rate is significantly increased by both, multi-interface communication (+25% data rate increase) as well as topology-awareness using pCAT (+17% throughput gain). For pure IEEE 802.11p-based transmission, the mean sensor data age is very high due to the relatively low coverage of the RSUs, resulting in high buffering durations. Here, the multi-interface approach is able to achieve significant benefits by using the cellular LTE link when no IEEE 802.11p connection is available.

## VI. Conclusion

In this paper, we presented an anticipatory multi-interface communication scheme that uses mobility prediction and channel quality estimation in order to optimize the transmission time of vehicular sensor data. Moreover, topology information about the surrounding buildings is taken into account for integrating the shadowing-behavior of the radio link into the transmission decision process.

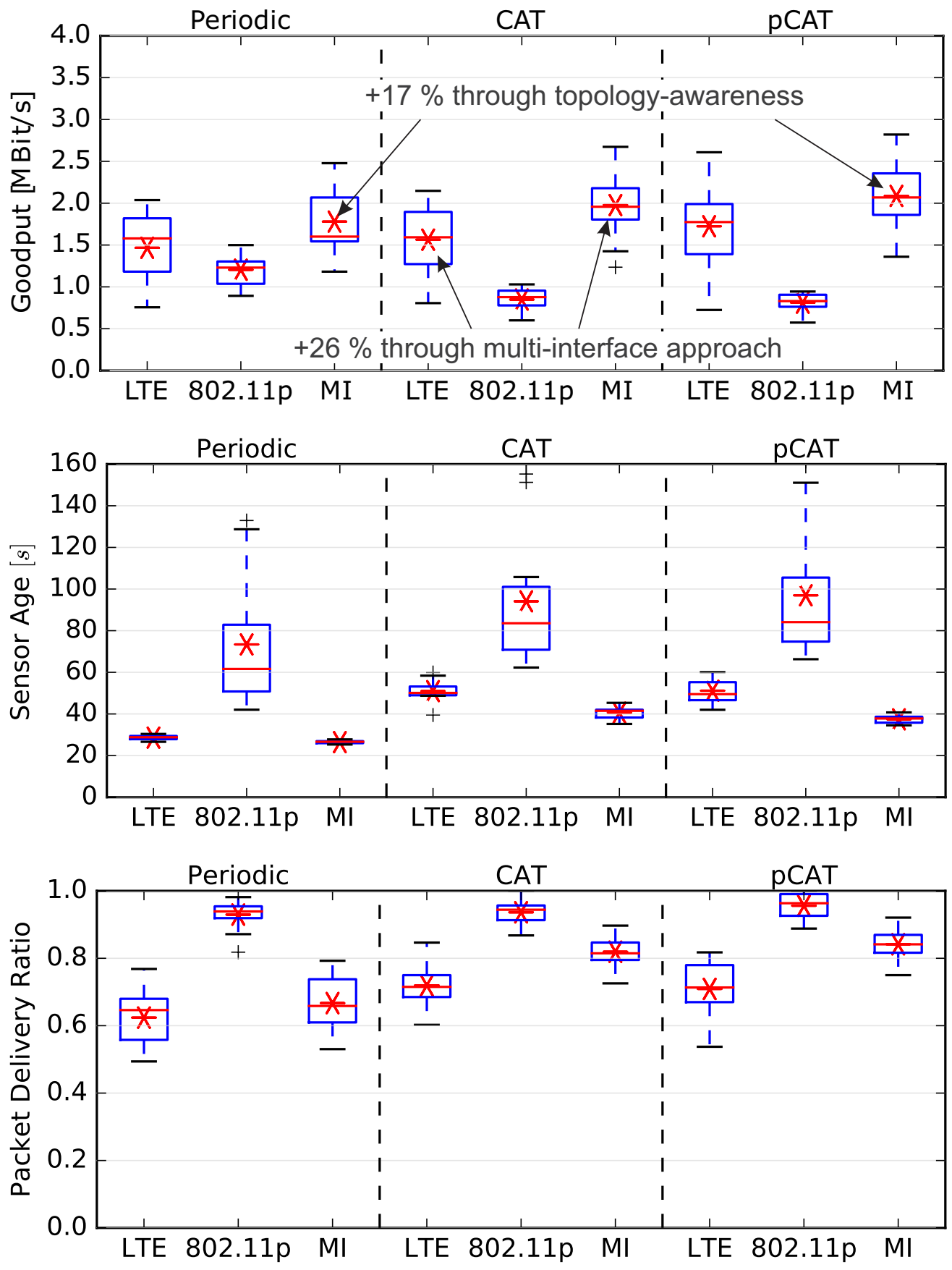

Fig. 12. Goodput and sensor data age for three different scenarios and transmission schemes. The tradeoff between goodput and data delay is reduced for the multi interface approach.

Using comprehensive simulations in OMNeT++, it was shown that the mean data rate and reliability of the car-to-cloud data transmissions can be significantly improved by mobility-prediction and multi-interface communication. In future work, the transmission scheme will be integrated into real-world devices for comprehensive field evaluation in the public cellular network. In addition, we will evaluate the use of network coding techniques to increase the robustness of the data transmissions.

## Acknowledgment

Part of the work on this paper has been supported by Deutsche Forschungsgemeinschaft (DFG) within the Collaborative Research Center SFB 876 "Providing Information by Resource-Constrained Analysis", project B4 and has been conducted within the AutoMat (Automotive Big Data Marketplace for Innovative Cross-sectorial Vehicle Data Services) project, which received funding from the European Union's Horizon 2020 (H2020) research and innovation programme under the Grant Agreement No 644657.